\documentclass[9pt]{article}
\usepackage{spconf,amsmath,graphicx}
\usepackage{amssymb,amsmath,amsfonts,eurosym,geometry,ulem,graphicx,caption,color,setspace,sectsty,comment,footmisc,caption,pdflscape,subfigure,array,hyperref}
\usepackage[utf8]{inputenc}
\usepackage{url}
\usepackage{multirow}
\usepackage{mathtools}
\newcommand{\mR}{\mathbb{R}}

\newcommand{\mc}{\mathcal}

\newcommand{\om}{\omega}

\newcommand{\Tha}{\Theta}

\newcommand{\lam}{\lambda}

\renewcommand{\leq}{\leqslant}
\renewcommand{\geq}{\geqslant}
\mathtoolsset{showonlyrefs}

\usepackage{algpseudocode}
\usepackage{algorithmicx}
\usepackage{algorithm}
\usepackage{caption}
\usepackage{eufrak}

\newtheorem{claim}{Claim}
\newtheorem{prop}{Proposition}
\newtheorem{Def}{Definition}
\newtheorem{assume}{Assumption}

\newtheorem{lemma}{Lemma}

\newcommand{\ket}{|\varphi\rangle}
\newcommand{\bra}{\langle\varphi|}
\newcommand{\ketprime}{|\varphi'\rangle}
\newcommand{\braprime}{\langle\varphi'|}

\newcommand{\keteta}{|\eta_i\rangle}
\newcommand{\ketetapp}{|\eta_{i'}\rangle}
\newcommand{\ketetaa}{|\eta_j\rangle}

\newcommand{\ketetaaa}{|\eta_k\rangle}

\newcommand{\braeta}{\langle\eta_i|}
\newcommand{\braetapp}{\langle\eta_{i'}|}

\newcommand{\braetaa}{\langle\eta_j|}
\newcommand{\braetaaa}{\langle\eta_k|}

\newcommand{\keti}{|\varphi_i\rangle}
\newcommand{\brai}{\langle\varphi_i|}
\newcommand{\ketk}{|k\rangle}
\newcommand{\brak}{\langle k|}

\newcommand{\ketj}{|\varphi_j\rangle} 
\newcommand{\braj}{\langle\varphi_j|}
\newcommand{\ketjp}{|\psi_{j}\rangle}
\newcommand{\brajp}{\langle\psi_{j}|}

\newcommand{\tr}{\text{Tr}}

\newenvironment{proof}[1][Proof]{\noindent\textbf{#1.} }{\ \rule{0.5em}{0.5em}}

\newcolumntype{L}[1]{>{\raggedright\let\newline\\arraybackslash\hspace{0pt}}m{#1}}
\newcolumntype{C}[1]{>{\centering\let\newline\\arraybackslash\hspace{0pt}}m{#1}}
\newcolumntype{R}[1]{>{\raggedleft\let\newline\\arraybackslash\hspace{0pt}}m{#1}}

\title{Quantum Man-in-the-middle Attacks: a Game-theoretic Approach with Applications to Radars}
\name{Yinan Hu and Quanyan Zhu}
\address{Tandon School of Engineering, New York University, USA}
\begin{document}
%\ninept
%
\maketitle
\begin{abstract}
The detection and discrimination of quantum states serve a crucial role in quantum signal processing, a discipline that studies methods and techniques to process signals that obey the quantum mechanics frameworks. However, just like classical detection, evasive behaviors also exist in quantum detection. In this paper, we formulate an adversarial quantum detection scenario where the detector is passive and does not know the quantum states have been distorted by an attacker. We compare the performance of a passive detector with the one of a non-adversarial detector to demonstrate how evasive behaviors can undermine the performance of quantum detection.
We use a case study of target detection with quantum radars to corroborate our analytical results. 
\end{abstract}
\begin{keywords}
Quantum detection, quantum man-in-the-middle attack, quantum radars, 
\end{keywords}
\section{Introduction}
\label{sec:intro}

The past two decades have witnessed a booming development of theories and applications of quantum systems. The discipline of quantum signal processing (QSP) \cite{eldar2002quantum_signal_processing} was founded under the construction of quantum mechanics frameworks \cite{von2018mathematical_QM} and studies the manipulation, detection and restoration of quantum signals.
One important branch of study of quantum systems is the detection and discrimination of states. In quantum signal processing, detecting the quantum state with good accuracy lays a foundation for many signal processing problems including the  reconstruction of the original quantum state. 

One challenge in designing quantum detection systems results from undesirable interactions \cite{sohbi2015decoherence_states} between the target mixed states and the exogenous environments.
In particular, the quantum states can be manipulated adversarially by malicious attackers, who undermine the detection performance significantly. For instance, a quantum man-in-the-middle attack \cite{fei2018quantum_mitm_qkd} jeopardizes the encryption process in quantum key distribution.
There is a need to formulate the adversarial manipulations of quantum states for quantum detection systems. 
    
In this paper, we develop a game-theoretical framework to study how a strategic attacker can undermine the performance of a quantum detector by distorting the quantum states produced by the quantum system. Game-theoretic frameworks have been widely implemented in studying attacker-defender relationships in cyber-security \cite{zhu2015game_CPS,huang2020dynamic_game_control}. In a generic quantum detection characterized by hypothesis testing scenario \cite{helstrom1976quantum_estimation_detection}, the detector (He) receives a sample of quantum states drawn from an unknown ensemble characterized by a density operator. The detector aims to make a decision regarding the genuine density operator that produces the quantum state. In our adversarial scenario, an attacker (She) observes the true density operator, intercepts the original quantum states, and sends strategically distorted quantum states to the `naive' detector, who applies the decision rule as if the received quantum states were untainted. We model the relationship between the attacker and the naive detector as a Stackelberg game \cite{Stackelberg2010market}, where the attacker plays the role of a leader and the detector the follower. 
    
Our analysis sheds light on fundamental limits on a quantum detector's performance in adversarial situations. We observe that by varying the threshold, one can depict the detection rate and false alarm rate of a naive quantum detector in terms of a receiver-operational-characteristic (ROC) curve \cite{helstrom1976quantum_estimation_detection}, which illustrates the detector performance. Compared to the curve for non-adversarial quantum detectors, the naive detector's performances deteriorate significantly as the attacker attenuate the component of the mixed state that is projected on the detection region.  Our results have clear implications for developing improved and attack-aware quantum detection systems that can combat strategically-designed quantum operations upon states. Our contributions are twofold: 
    
The rest of the paper is organized as follows. In Section \ref{sec:format}, we formulate man-in-the-middle-attack as a Stackelberg game and compute his optimal strategies. In Section \ref{sec:quantum_radar}, we illustrate our formulation with a case study in target detection using quantum radars of a particular type. We conclude in Section \ref{sec:conclusion}.
\paragraph*{Notations}
We denote $\mc{H}$ as the Hilbert space (over the set of real numbers $\mR$) and $\mc{H}^*$ as its dual space.  We inherit Dirac's bra-ket notations  \cite{dirac1981principles} to denote generic quantum states: $\bra\in\mc{H}^*,\ket\in\mc{H}$. Let $B(\mc{H})$ be the space of all positive, Hermitian and bounded operators from $\mc{H}$ to itself. Let $\mc{S}$ be the subset of $B(\mc{H})$ such trace of its operators is $1$.  
%A generic density operator upon $\mc{H}$ is denoted as $\rho$. Thus we could specify the space $\mc{S}$ as 
%\begin{equation}
%    \mc{S} = \{\rho\in B(\mc{H}):\; \tr(\rho) = 1,\;\tr(\rho^2)\leq 1\}.
%\end{equation} 
In addition, we denote $\mc{V}$ as the space of projection-valued measurements \cite{von2018mathematical_QM}.
We denote $\mathbf{1}\in B(\mc{H})$ as the identity operator. For any operator $A\in B(\mc{H})$, we denote its conjugate transpose as $A^{\dagger}$.

\section{The formulation of a quantum man-in-the-middle (MITM) attacker}
\label{sec:format}
We begin by considering a non-adversarial detection formulation based on quantum hypothesis testing introduced in \cite{helstrom1976quantum_estimation_detection}. Suppose that there is an unknown target quantum system characterized by a density operator $\rho\in\mc{S}(\mc{H})$. We consider the binary hypothesis testing scenario, assuming that the nature of the system specifies two possible choices for $\rho=\rho_0$ or $\rho=\rho_1$, each of which forms a hypothesis $H_0,H_1$ as follows: 
\begin{equation}
    H_0: \rho_0= \sum_{j = 1}^d{r^0_j\ketjp\brajp},\;\;H_1: \rho_1 = \sum_{i = 1}^d{r^1_i\keti\brai}, \label{hypo_rho1_rho0}
\end{equation}
with $\sum_{i}{r^1_i} = \sum_{j}{r^1_j} = 1,\;r^1_j,r^0_i\geq 0$ and $\ketjp,\keti\in \mc{H},\; i,j=1,2,\dots,d$. We assume that $\{\ketjp\},\{\keti\}$ span the Hilbert space $\mc{H}$ of dimension $d$.  Notice that we do not assume that $\rho_0,\rho_1$ commute; i.e. $\{\ketjp\}^d_{j=1},\{\keti\}^d_{i=1}$ may be two different bases of $\mc{H}$. The target system produces quantum states $\ket\in\mc{H}$ which are collected by a detector (He), who wants to arrive at a decision rule $\delta\in \mc{H}\rightarrow [0,1]$ on which is the genuine characterization of the target system in \eqref{hypo_rho1_rho0}: $\delta(\varphi) = k,\;k\in\{0,1\},$ when he thinks the hypothesis $H_k$ holds true. According to the measurement postulate of quantum mechanics \cite{von2018mathematical_QM}, the detector decides by applying a measurement $\Pi_1\in\mc{V}$ upon the received quantum state $\ket$: $\delta(\varphi) = \bra \Pi_1 \ket$. 
The decision rule $\delta$, or equivalently the measurement $\Pi_1$, leads to a detection rate $P_D: \mc{V}\rightarrow [0,1]$  and a false alarm rate $P_F:\mc{V}\rightarrow [0,1]$ as follows:
\begin{align}
     P_D(\Pi_1) = \tr(\Pi_1\rho_1), \;\; P_F(\Pi_1) = \tr(\Pi_1\rho_0).
     \label{PD_PF}
\end{align}
In Bayesian hypothesis testing formulation, the detector knows the probabilities that $H_1,H_0$ hold true are $c_1,\;c_0$ ($c_1 + c_0 = 1$) respectively. Referring to \cite{helstrom1976quantum_estimation_detection} and denoting $\tau = \frac{c_1}{c_0}$, detector's optimal solution measurement $\Pi^*_1$ can be designed as follows:
\begin{equation}
    \Pi^*_1(\tau) = \sum_{\eta_j>0}{\ketetaa\braetaa},
    \label{sol_stackelberg_detector}
\end{equation}
where $\{\ketetaa\}^d_{j=1}$ are orthogonal eigenstates of $\rho_1-\tau\rho_0$ with eigenvalues $\eta_j$.
\begin{figure}
    \centering
    \includegraphics[scale=0.3]{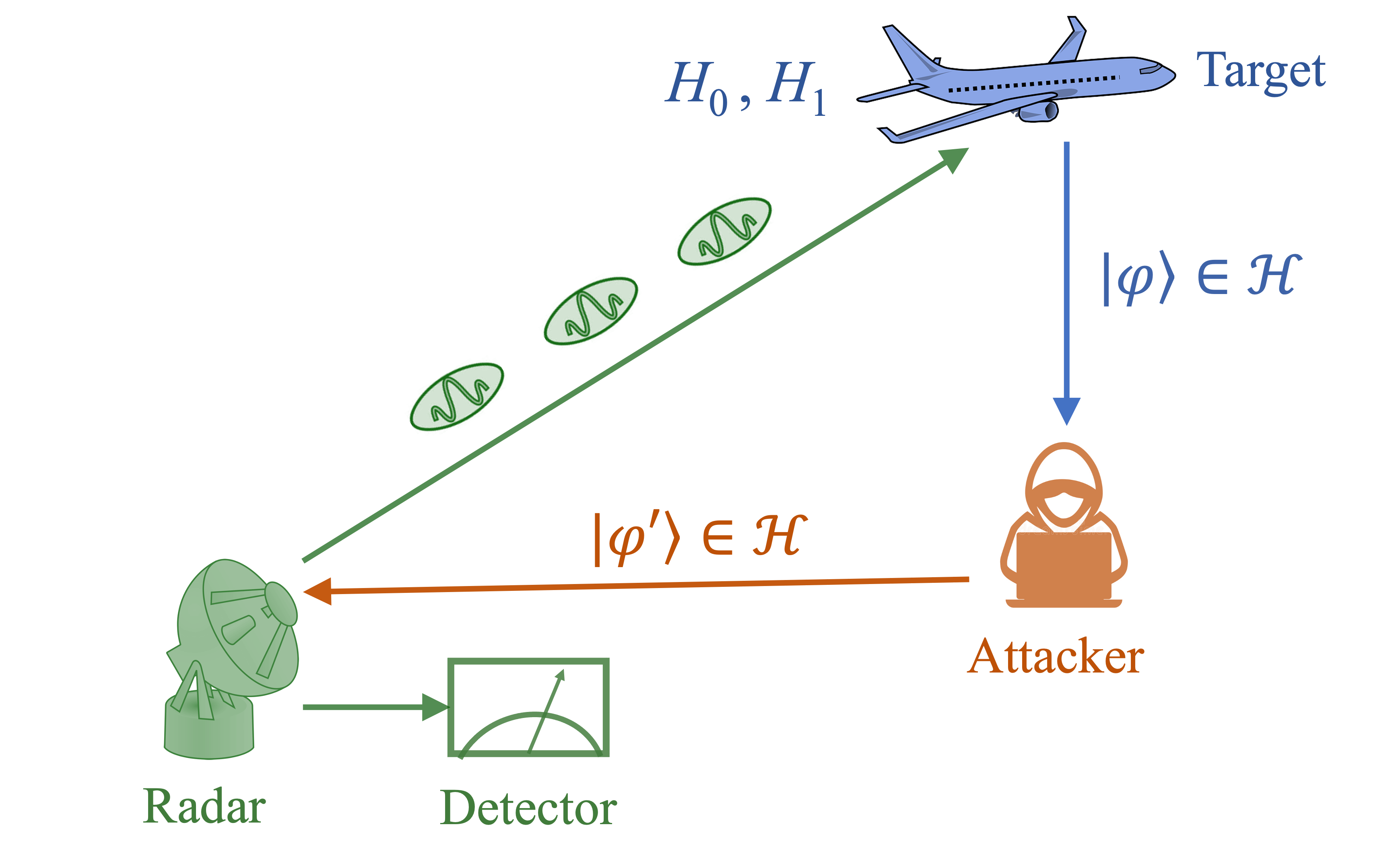}
    \caption{The scheme of quantum state detection with adversaries. The target system produces quantum states $\ket$ from different ensembles depending on the true hypothesis $H_0,H_1$. The attacker hijacks and replaces the clean quantum states with the distorted ones. One case study is the target detection of quantum radars \cite{slepyan2021quantum_radar_lidar}.}
    \label{fig:quantum_man_in_the_middle}
\end{figure}
\subsection{Quantum man-in-the-middle attack: a Stackelberg game formulation}

We now introduce the scenario (as in Figure \ref{fig:quantum_man_in_the_middle}) of adversarial quantum detection in which an attacker (She) stands between the target system and detector, intercepts the quantum state $\ket\in\mc{H}$ from the target systems and sends out strategically distorted signal $\ketprime\in \mc{H}$ to the detector to undermine his performance.  
We assume that the detector is passive and `naive' about the distortion and designs his optimal measurement operation as if the quantum state $\ketprime$ were directly generated from the target system.   Upon receiving $\ketprime$, the detector designs an optimal decision rule $\hat{\delta}^*:\mc{H}\rightarrow [0,1]$ by measuring the quantum state $\ketprime$ from attacker with measurement, denoted as $\Pi_1$ as follows: $$\hat{\delta}^*(\varphi') = \braprime \Pi^*_1 \ketprime.$$ 
Thus we let the detector's cost function $u_B:\mc{V}\times \mc{S}\times \mc{S}\rightarrow \mR$ be the Bayesian risk, which is a weight sum of \textit{counterfactual} miss rate and \textit{counterfactual} false alarm rate in \eqref{PD_PF} as if the quantum state were untainted:    
\begin{equation}
\begin{aligned}
    \underset{\rho'_1,\rho'_0\in \mc{S}}{\min}&u_B(\Pi_1,\rho'_1,\rho'_0) \\
    \Longleftrightarrow  \underset{\rho'_1,\rho'_0\in \mc{S}}{\min}& c_1 \tr((\mathbf{1}-\Pi_1)\rho_1) + c_0\tr(\Pi_1\rho_0),\;\;  
\label{detector_stackelberg_passive}
\end{aligned}
\end{equation}
which leads to his optimal measurement $\Pi^*_1$ as in \eqref{sol_stackelberg_detector}.
The attacker, after observing the true hypothesis ($H_1$ or $H_0$) on the target systems, obtains the quantum state $\ket\in\mc{H}$. Based on his observation on the true density state ($\rho_1$ or $\rho_0$), the attacker designs quantum noisy operations ${E}^1,E^0\in B(\mc{H})$ acting upon the received state $\ket$ to create a distorted quantum state $\ketprime$.  According to \cite{nielsen2002quantum_information}, we formulate the resulting noisy density operators $E^0, E^1$  as the `operator-sum' representation as follows:
\begin{equation}
\begin{aligned}
 \mc{E} &= \{(E^0, E^1)\in {B}(\mc{H})^{\otimes 2}:   \sum_{k}{E^l_kE^{l\dagger}_k} = \mathbf{1},\;\;l\in\{0,1\}\}.
 \label{action_space_attacker_Alice}
\end{aligned}
\end{equation}
\begin{comment}
Equivalently, we could consider that the attacker manipulates the true density state $\rho$ based on the true hypothesis $H_0,H_1$ in the following way: $E^l(\rho) = \sum_{k}{E^l_k\rho E^{l\dagger}_k},\;l=0,1.$
\end{comment}
We could treat the space $\mc{E}$ in \eqref{action_space_attacker_Alice} as the attacker's action space. Yet we argue through the following lemma that it is equivalent to characterize the attacker's strategies as a pair of density operators $\rho'_1,\rho'_0$. 
\begin{lemma}[Equivalency of attacker's strategies]
\label{lemma:equiv_attacker_action}
Let $\rho_1,\rho_0$ be the two density operators in \eqref{hypo_rho1_rho0}. Then, for any $\rho'_1,\rho'_0\in\mc{S}$ there exist operations ${E}^1,{E}^0\in \mc{E}$ of the operator-sum representation such that ${E}^1(\rho_1) = \rho'_1,\;{E}^0(\rho_0) = \rho'_0$.
\label{lemma:existence_density_op}
\end{lemma}

Knowing detector's strategy \eqref{sol_stackelberg_detector} the detector designs her optimal strategies $\rho'^*_1,\rho'^*_0$ by minimizing her utility function $u_A: \mc{S}\times \mc{S}\times \mc{V}\rightarrow \mR$ as follows:
\begin{equation}
\begin{aligned}
\underset{\rho'_0,\rho'_1\in\mc{S}}{\min}&\;{u_A(\rho'_1,\rho'_0,\Pi^*_1) }, \\
\Longleftrightarrow\underset{\rho'_0,\rho'_1\in\mc{S}}{\min}&\;\tr(\Pi^*_1\rho'_1) + \lambda [S(\rho'_1\|\rho_1) +  S(\rho'_0\|\rho_0)], 
\end{aligned}
\label{attacker_stackelberg}
\end{equation}
where we adopt the Von-Neumann relative entropy \cite{nielsen2002quantum_information} for any two density operators $\nu_1,\nu_0\in\mc{S}$ as
$S(\nu_1\|\nu_0) := \tr(\nu_1(\ln \nu_1-\ln\nu_0))
$, and $\Pi^*_1$ is the solution to detector's optimization problem \eqref{detector_stackelberg_passive}. 

In the objective function of \eqref{attacker_stackelberg}, the attacker trades off between undermining the genuine detection rate (the first term) and minimizing the loss incurred from distorting quantum states through noisy gates (the second term). Such a loss term originates from the detector's awareness of the distortion if the received quantum states deviate significantly from the `normal' ones. 
The parameter $\lambda\in \mR_+$ is a regularization parameter characterizing the attacker's intentions.

Summarizing the formulations raised in \eqref{detector_stackelberg_passive} and \eqref{attacker_stackelberg}, we arrive at the game between passive quantum detector and the attacker $\mc{G}$, which is a Stackelberg game \cite{Stackelberg2010market} defined as follows.
\begin{Def}
We define the relationship between the passive quantum detector and the attacker as a Stackelberg game $\mc{G}$ with the following tuples
\begin{equation}
    \mc{G} = \langle \mc{I}, \Tha, \mc{H}, \mc{E}, \mc{V}, u_A, u_B \rangle,
\end{equation}
where $\mc{I}= \{\text{Attacker},\;\text{Detector}\}$ refers to the set of players; $\Tha = \{H_0,H_1\}$ refers to the set of true hypotheses specified in \eqref{hypo_rho1_rho0}; $\mc{H}$ is the Hilbert space; the set $\mc{E}$ in \eqref{action_space_attacker_Alice} characterizes the attacker's strategy space; the set of measurements $\mc{V}$ characterizes the detector's space of decision rules. The functions $u_A, u_B$ specified in \eqref{detector_stackelberg_passive} and \eqref{attacker_stackelberg} the objectives of the attacker and the detector, respectively. 
\label{def:stackelberg_detector}
\end{Def}
We now state the attacker's optimal design of distorted mixed densities $\rho'^{*}_1,\rho'^{*}_0$ into the following proposition:
\begin{prop}[Attacker's optimal strategies]
\label{attacker_strategies_stackelberg_prop}
Let $\mc{G}$ be the Stackelberg game between the attacker and the detector (the passive quantum detector) as mentioned in definition \ref{def:stackelberg_detector}. Then the attacker's optimal strategies $\rho'^{*}_1,\rho'^{*}_0$ are expressed as follows: 
\begin{align}
    \rho'^*_0 &= \rho_0,
     \label{sol0_attacker_Stackelberg} \\
     \rho'^*_1 &=  \frac{\exp(\ln\rho_1 - \frac{1}{\lambda}\Pi^*_1) }{\tr(\exp(\ln\rho_1 - \frac{1}{\lambda}\Pi^*_1))}.
    %\rho'^*_1 &=  \frac{\sum_{i=1}^d{r^1_i e^{-\frac{\beta_i}{\lambda}}\keti\brai}}{\sum_{i=1}^d{r^1_i e^{-\frac{\beta_i}{\lambda}}}},
    \label{sol1_attacker_Stackelberg}
\end{align}
%with 
%\begin{equation}
%    \beta_i = \sum_{\eta_j>0}{|\langle\varphi_i|\eta_j\rangle|^2}.
%    \label{beta_i}
%\end{equation}
\end{prop}

\begin{comment}
\begin{proof}
To make sure the $S(\rho'_1,\rho_1) < \infty$, we know the optimal strategy $\rho^*_1$ must have the same `kernels' as $\rho_1$ as follows:
\begin{equation}
    \rho'^{*}_1 = \sum_{i=1}^d{s_i \keti\brai},\;\;\sum_{i=1}^d{s_i} = 1.
\end{equation}
Both terms in the objective functions involves trace operations, which are independent of the basis to be chosen. We compute the first trace term through the basis $\{\keteta\}$ from \eqref{sol_stackelberg_detector} and the second via the basis $\{\keti\}$ from \eqref{hypo_rho1_rho0}. We can rewrite 
\begin{equation}
    u_B(\Pi_1,\rho'_1,\rho'_0) = \sum_{i=1}^d{s_i\beta_i} + \lambda \sum_{i=1}^d{s_i\ln\frac{s_i}{r^1_i}},
\end{equation}
with 
\begin{equation}
    \beta_i = \sum_{\eta_j>0}{|\langle\varphi_i|\eta_j\rangle|^2}.
    \label{beta_i}
\end{equation}
Applying the first-order condition of $u_B$ in terms of $s_i,\;i=1,2,\dots, d$ and 
Considering the normalization requirement $\tr(\rho'^*_1)=1$ yields the conclusion in the proposition.
\end{proof}
\end{comment}

We have several remarks on the attacker's and the detector's optimal strategies. First, as $\lambda \rightarrow 0$, the penalty for distorting the mixed state vanishes, and the attacker manages to suppress all the components in $\Pi_1$. 
On the other hand, when $\lambda\rightarrow \infty$, the attacker's optimal strategies $\rho'^*_1 = \rho_1$, meaning the attacker does not distort the original density operator at all because the penalty for distorting the mixed states becomes infinitely high.  

Secondly, we can interpret the attacker's optimal manipulation of mixed density states as in Proposition \eqref{attacker_strategies_stackelberg_prop} as the attenuation of the components of the original quantum state $\rho_1$ upon the subspace spanned by the base states $\{\keteta\}_i$ lying in the region of detection. Applying Baker-Campbell-Hausdorf formula to the nominator of the RHS, of \eqref{sol1_attacker_Stackelberg} we obtain 
\begin{equation}
    \rho'^*_1 = \frac{\rho_1e^{-\frac{1}{\lambda}\Pi^*_1}e^{-\frac{1}{2\lambda}[\ln\rho_1,\Pi^*_1]+\dots}}{\tr(\exp(\ln\rho_1 - \frac{1}{\lambda}\Pi^*_1))},
\end{equation}
where ``$\dots$" refers to additional terms involving iterative Poisson brackets of $\ln\rho_1$ and $\frac{1}{\lambda}\Pi_1$. The product $\rho_1e^{-\frac{1}{\lambda}\Pi^*_1}$ indicates the attenuation of the original mixed state $\rho_1$ on the subspace controlled by the parameter $\frac{1}{\lambda}$. The the remaining factor $e^{-\frac{1}{2\lambda}[\ln\rho_1,\Pi_1]+\dots}$, where the commutator $[\ln\rho_1,\Pi_1]$ is included in the exponent, implies that the quantum uncertainty principle also affects attacker's optimal strategies.

Thirdly, when $\rho_1,\Pi^*_1$ commute (a sufficient condition would be $\rho_1,\rho_0$ commute), the attacker's optimal strategies \eqref{sol0_attacker_Stackelberg}\eqref{sol1_attacker_Stackelberg} reduce to the strategies in classical Stackelberg hypothesis testing games as formulated in Section II of \cite{hu2022game_NP}.

%If we denote the subspace spanned by $\{\ketetaa\}_{\eta_j>0}$ as $\mc{M}$ and its orthogonal complement $\mc{M}^{\perp}$, then $\mc{H} = \mc{M}\oplus \mc{M}^{\perp}$, and the detector will decide $H_1$ once the received state is $\ket\in\mc{M}$. %The detection rate of the detector is reduced because the attacker manipulates the quantum states by `projecting' them onto $\mc{M}^{\perp}$. 

We can now compute the \textit{genuine} detection rate $\bar{P}_D: \mc{V}\rightarrow [0,1]$ under the attacker's manipulation:
\begin{equation}
    \bar{P}_D(\Pi^*_1) = \tr(\Pi^*_1\rho'^{*}_1), \; \bar{P}_F(\Pi^*_1) = \tr(\Pi^*_1\rho'^{*}_0),
    \label{bar_PD}
\end{equation}
where $\rho'^*_1$ is obtained from \eqref{sol1_attacker_Stackelberg}. We have the following statement:
\begin{prop}
\label{prop:fundamental_limit_detection_rate}
Let $(\rho'^*_0,\rho'^*_1,\Pi^*_1)$ be the optimal strategy profile for the Stackelberg game $\mc{G}$ derived in \eqref{sol0_attacker_Stackelberg}\eqref{sol1_attacker_Stackelberg} and \eqref{detector_stackelberg_passive}, respectively. Recall $P_D,\bar{P}_D$ as the counterfactual and genuine detection rates defined in \eqref{PD_PF} and \eqref{bar_PD}, respectively. Then, under some technical assumptions we have:
\begin{equation}
    \;P_D(\Pi^*_1)e^{-\frac{1}{\lambda}}\leq \bar{P}_D(\Pi^*_1)\leq P_D(\Pi^*_1),\;\forall \Pi^*_1\in\mc{V}.
\end{equation}
\end{prop}
Proposition \ref{prop:fundamental_limit_detection_rate} characterizes an upper and lower bound of the genuine detection rate under strategically designed quantum operations. 
%where $\beta_i,\;i\in[d]$ is obtained in \eqref{beta_i}.
\section{Case study: target detection using quantum radars}
\label{sec:quantum_radar}
We now apply the formulation of the quantum man-in-the-middle attack  discussed in Section \ref{sec:format} to study spoofing in quantum radar detection. A quantum radar is a standoff detection system using photons to explore some quantum phenomena to strengthen its capacity to detect targets of interest. In this section, we assume that our quantum radar generates non-entangled, monochromatic, coherent \cite{torrome2020quantum_radar} photonic quantum states subject to noise in line with Llyod's theory \cite{lloyd2008quantum_illumination}.

We illustrate the MITM attack scheme in Figure \ref{fig:quantum_man_in_the_middle}: a quantum radar (detector) determines on the absence or presence of the target object based on the reflective photon-based quantum states. An attacker blocks the transmission of reflective quantum signals, manipulates them through noisy quantum gates selected from the action space in \eqref{action_space_attacker_Alice}, which can be implemented using photonic quantum circuits \cite{obrien2007optical_quantum_computing}, and sends out manipulated quantum signals. 
%There are multiple schemes implementing the quantum radars, depending on whether quantum states are coherent and entangled or not. 
 We use the mean number representation to characterize photonic quantum states from the reflective signals \cite{fox2006quantum_optics}. 
We associate the hypotheses $H_0,H_1$ with the mixed state of reflective signals under the absence $H_0$ or presence $H_1$ of the target object in Figure \ref{fig:quantum_man_in_the_middle}, respectively as follows:
\begin{equation}
\label{H_0_H1_target_no_target}
   \begin{aligned}
   H_0:\rho_0 &= (1-N_B)|0\rangle\langle 0| + N_B{\ketk\brak}, 
   \\
    H_1: \rho_1 &= (1-x)\left((1-N_B)|0\rangle\langle 0| + N_B{\ketk\brak}\right) + x|l\rangle\langle l|,
\end{aligned} 
\end{equation}
where $x\in[0,1]$ refers to the reflective index, and $N_B\in[0,1]$ characterizes the noise of the environment. 
\begin{comment}
Depending on the entanglement of photons, there are four types of quantum radars: type 1 sensors transmit un-entangled non-coherent quantum states of light; type 2 sensors transmit coherent light (classical states of light), yet uses quantum photonic sensors to enhance detection performance; type 3 transmit entangled quantum states of light with the receiver, and the signal beams are not entangled with itself; recently, there is a type 4 quantum sensor that produces entangled quantum light states with multiple entangled photons.  We adopt the type I quantum sensors, in which there is no entanglement between photons. The formulations that we developed in previous sections can be generalized to other quantum light states as well. 
\end{comment}
\begin{comment}
In the meantime, we assume that the prior belief of the true hypothesis $H_0, H_1$ is $p,1-p$ respectively. The two underlying density operators $\rho_0,\rho_1$ under each hypothesis, the prior belief, the reflect index and $N_B$ are all common knowledge between the spoofer and the detector.
In a Stackelberg game formulation as indicated in definition \ref{def:stackelberg_detector}, the spoofer (leader, attacker) intercepts the original signal and wants to produce a quantum state that is difficult for the receiver (follower, detector, detector) to discriminate. 
\end{comment}

\begin{figure}
    \centering
    \includegraphics[scale=0.12]{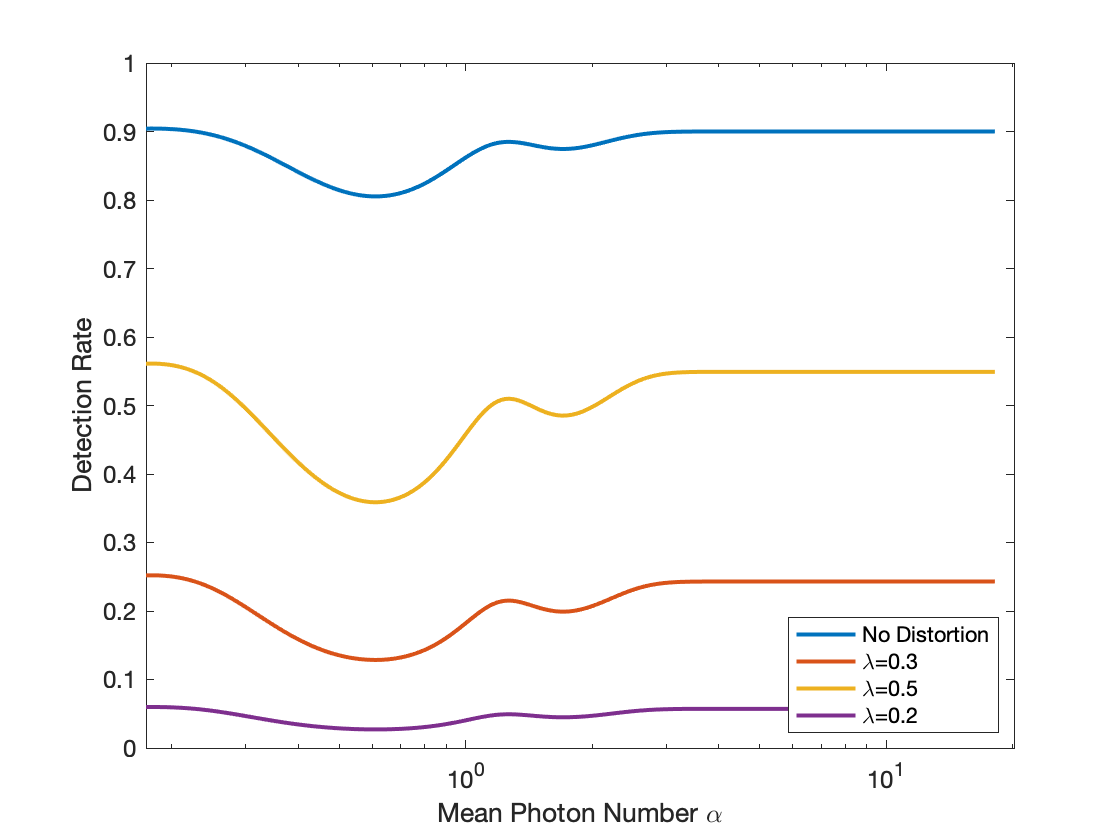}
    \caption{The genuine detection rate $\bar{P}_D$ in \eqref{bar_PD} in terms of the mean photon number of the given mixed state $\rho_1$ in \eqref{H_0_H1_target_no_target} under different attacks. We fix the detector's threshold $\tau = 1$, $N_B,k,x$ and let $l$ vary. The genuine detection rates become low when the mean photon number is close to $k$, suggesting the detector performs the worst when the reflective signal resembles the noise.}
    \label{fig:PD_mean_photon}
\end{figure}

\begin{figure}
    \centering
    \includegraphics[scale=0.26]{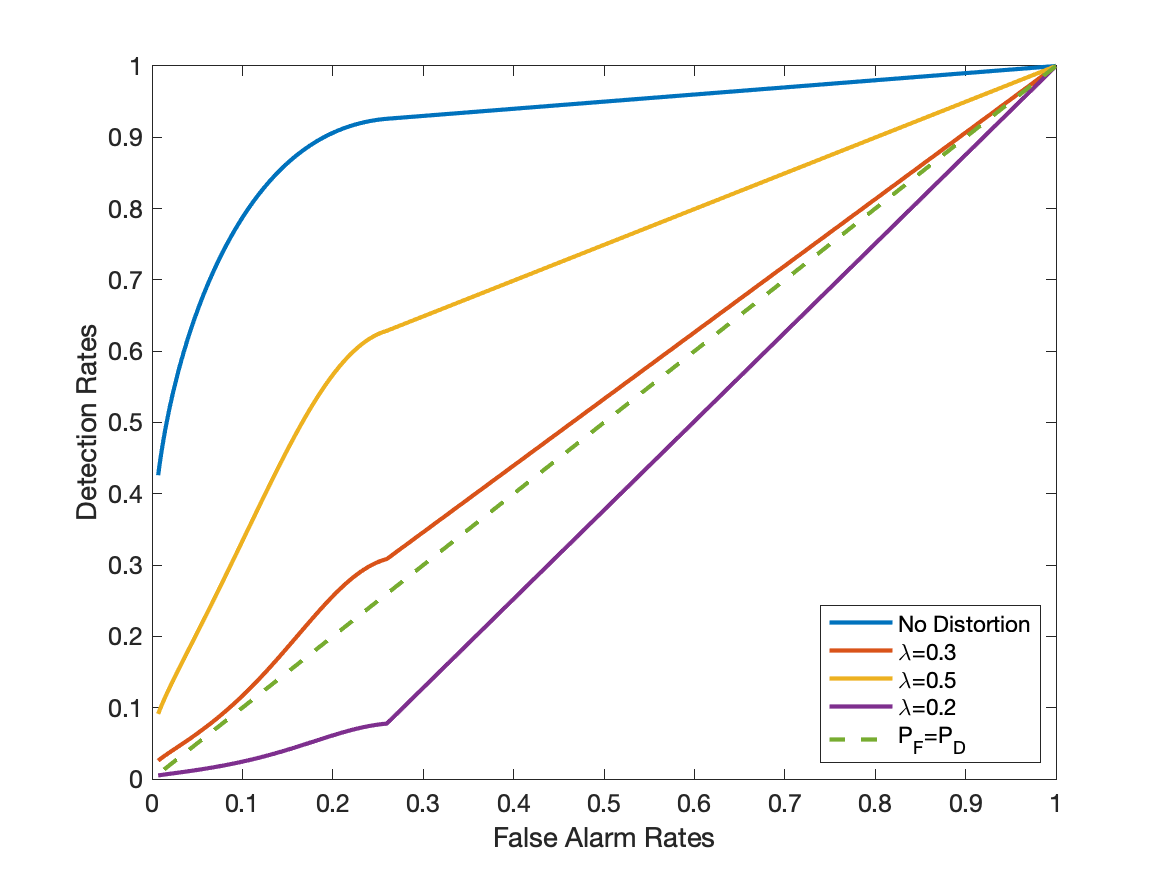}
    \caption{The ROC curves of the naive quantum detector under distorted state $\rho'^*_1$  parameterized by different values of parameter $\lambda$ in \eqref{sol1_attacker_Stackelberg}.  We choose intrinsic coefficients of the detector to be $N_B=0.4, l=2, k=1,x=0.9$ in \eqref{H_0_H1_target_no_target}. }
    \label{fig:roc_stackelberg}
\end{figure}

Based on the true hypothesis $H_0,H_1$ and the input quantum state, the spoofer/attacker produces distorted quantum mixed densities $\rho'^{*}_0,\rho'^{*}_1$ as in \eqref{attacker_stackelberg}. We choose $N_B = 0.4, l=2, k=1, x= 0.9$  in \eqref{H_0_H1_target_no_target} and depict in Figure \ref{fig:PD_mean_photon} the relationship between the detection rate in terms the mean number of photons under attacks parameterized by different choices of $\lambda$. We observe that no matter the distortion level, the quantum detector reaches its worst detection rates when the mean photon number is around $0.62$, and its second worst rate is at $1.69$. Nevertheless, different distortion levels cause a contrast in detection rates among all mean photon number states. In Figure \ref{fig:roc_stackelberg}, we plot the receiver-operational-characteristic (ROC) curves of classical quantum detectors and passive quantum detectors when the attacker's manipulation strategy is controlled under different choices of $\lambda$. We observe that the passive quantum detectors perform worse than the classical quantum detectors because the spoofer manipulate the states to undermine the detection rates. Lower values of $\lambda$ lead to stronger distortion of states and thus worse performance in terms of the ROC curves.   

\section{Conclusion}
\label{sec:conclusion}
We have formulated the quantum man-in-the-middle attack under the Stackelberg framework to study the fundamental limit of the reduction of the detection performance. We characterize the attacker's optimal strategies as a suppression of the target's density operator on the region of detection, which is reshaped by the quantum effects. We show that a passive quantum detector suffers an exponential reduction of detection rate in the worst case. Our numerical results also imply how adversarial attacks affect the performance of quantum radars. 

One direction to extend our adversarial quantum detection framework would be to consider various detection frameworks such as side-channel attack detection \cite{lemarchand2020em_side_channel_attack}. 
It would be also possible to consider hybrid detection models taking advantage of both classical and quantum information as inputs.

%\vfill
%\pagebreak
\newpage
\bibliographystyle{IEEEbib}
\bibliography{rl}
\newpage
\section{Appendix}
\subsection{Proof of Lemma \ref{lemma:existence_density_op}}
\begin{proof}
It suffices to prove that $\rho'_1$ and $E_1$ are equivalent in characterizing the attacker's generic strategies when observing the hypothesis $H_1$. Without loss of generality we assume further that $\rho_1$ is strictly positive, that is $r^1_i>0$ for all $i$. We prove two claims as follows.
\begin{claim}
For every $\rho'_1\in\mc{S}$, there always exist a collection of operators $\{E^1_1,E^1_2,\dots,E^1_{d^2}\}$ such that 
\begin{equation}
    \rho'_1 = E^1(\rho_1) = \sum_{k=1}^{d^2}{E^1_k \rho_1 E^{1\dagger}_k}.
    \label{rho1_prime}
\end{equation}
\end{claim}
\textbf{Proof of Claim 1.} Since we assume $\text{dim}(\mc{H}) = d$, we know the distorted density operator $\rho'_1$ has $d^2$ degrees of freedom. 
Since $\rho'_1\in\mc{S}$, we can parameterize them in the basis $\keti$ as well with the coefficients $a'_{ij}$. 
We can express them in terms of the basis $\{\ketj\}^d_1$ as follows:
\begin{equation}
\rho'_1 = \sum_{i,j=1}^{d}{a'_{ij}\keti\braj}.
\label{eq:rho1p_rho0p}
\end{equation}
The impact of the quantum operations upon $\rho_1$can be characterized by the change of coordinates of $\rho_1$ into the ones of $\rho'_1$ under the basis $\keti\braj$.
Now the operations $\{E^1_k\}$ turn the coordinates from $a_{ij}$ into $a'_{ij}$, respectively. Under the representation (or basis) of $\{\ketj\}^d_1$, we can select the matrix representation of $E^1_k,E^0_k$, where $k=1,2,\dots,d^2$ as follows:
\begin{equation}
\begin{aligned}
 E^1_k &= \begin{bmatrix}
    0 & \hdots & \hdots & \hdots &  0 \\
    0 & \hdots & \sqrt{\frac{a'_{ij}}{a_{ij}}} & \hdots & 0\\
     \vdots & \vdots & \ddots &\vdots  & \vdots  \\
    0 & \hdots & \hdots & \hdots &  0
    \end{bmatrix}.\;
    \label{eq:matrix_E1_k}
\end{aligned}
\end{equation}
$(E^1_k)_{i(k),j(k)} = \sqrt{\frac{a'_{ij}}{a_{ij}}}\;$
where $i(k),j(k)$ refers to the row index and column index corresponding to the subscript $k$. The rest of the entries for $E^1_k$ are all zero.
Then it is clear to verify that 
\begin{equation}
    E^1({\rho_1}) = \rho'_1
\end{equation}
as claimed. 

Also, we have the following conclusion.
\begin{claim}
Let $\rho_1\in S$ be a density operator. For every $E^1\in B(\mc{H})$, we can find a $\rho'_1\in\mc{S}$ such that 
    \begin{equation}
        \rho'_1 = \sum_{k=1}^{d^2}{E^1_k\rho_1E^{1\dagger}_k}.
        \label{rho1_to_rho1prime}
    \end{equation}
\end{claim}
\textbf{Proof of claim 2.} Our goal is to prove that $\rho'_1$ obtained in \eqref{rho1_to_rho1prime} meets the requirements of a density operator, that is, it has a trace of $1$; it is positive definite; it is symmetric (or Hermitian).  We first of all notice 
\begin{equation}
\begin{aligned}
    \tr(\rho'_1) &= \tr\left(\sum_{k=1}^{d^2}{E^1_k\rho_1 E^{1\dagger}_{k}}\right) = \sum_{k=1}^{d^2}\tr(E^1_k\rho_1 E^{1\dagger}_{k}) \\
    & = \sum_{k=1}^{d^2}\tr( E^{1\dagger}_{k}E^1_k\rho_1) = \tr\left(\sum_{k=1}^{d^2}{ E^{1\dagger}_{k}}E^1_k\rho_1\right) \\
    & = \tr[\left(\sum_{k=1}^{d^2}{ E^{1\dagger}_{k}}E^1_k\right)\rho_1]
\end{aligned}  
\end{equation}
Noticing that $\sum_{k=1}^{d^2}{ E^{1\dagger}_{k}}E^1_k = \mathbf{1}$, we conclude 
\begin{equation}
    \tr(\rho'_1) = \tr(\rho_1) = 1.
\end{equation}
Next we prove symmetry of $\rho'_1$. For convenience we denote $\rho_{ij},\rho_{ij}$ as the matrix entries of $\rho_1,\rho'_1$ respectively. It suffices to assume that every matrix $E^1_k$ is written in the form similar as the one in \eqref{eq:matrix_E1_k}.  
We want to show that 
\begin{equation}
    \om(j,k) = E_{i_1j_1} \rho_1 E^{\dagger}_{i_2j_2} + E_{i_2j_2}  \rho_1E^{\dagger}_{i_1j_1} 
    \label{eq:omega}
\end{equation}
is also symmetric for all choices of $i_1,i_2,j_1,j_2\leq d$. Here $E_{i_1j_1}$ is a $d$ by $d$ matrix where every entry is zero except $(i_1,j_1)$. By computing entry-wisely, we notice that 
\begin{equation}
    [E_{i_1j_1} \rho_1 E^{\dagger}_{i_2j_2}]_{kl} = \begin{cases}
    a_{i_1j_1}\rho_{i_1i_2}a_{i_2j_2} & \text{if}\;k=i_1,\;k=i_2, \\
    0 & \mbox{else}
    \end{cases}
    \label{eq:matrix_E1_k2}
\end{equation}
Similarly,
\begin{equation}
    [E_{i_2j_2} \rho_1 E^{\dagger}_{i_1j_1}]_{kl} = \begin{cases}
    a_{i_2j_2}\rho_{i_2i_1}a_{i_1j_1} & \text{if}\;k=i_2,\;k=i_1, \\
    0 & \mbox{else}.
    \end{cases}
    \label{eq:matrix_E1_k3}
\end{equation}
Substituting \eqref{eq:matrix_E1_k2} and \eqref{eq:matrix_E1_k3} into \eqref{eq:omega} and considering $\rho_{i_1i_2} = \rho_{i_2i_1}$ due to symmetry of $\rho_1$, we get $\om(j,k)$ is indeed symmetric. Since every matrix can be written as a linear combination of $\{E_j\}_{1\leq j\leq d^2}$, we get that $E^1_k\rho_1E^1_k$ is symmetric for all $1\leq k\leq d^2$. As a consequence, $\rho'_1$ as expressed in \eqref{rho1_prime} is symmetric too. 

\begin{comment}
Since $\rho_1 = \rho^T_1$, we have 
\begin{equation}
\begin{aligned}
    &\rho'^{T}_1 = (\sum_{k=1}^{d^2}{E^1_k\rho_1E^{1\dagger}_k})^T = \sum_{k=1}^{d^2}{(E^1_k\rho_1E^{1\dagger}_k)^T}\\
    &=\sum_{k=1}^{d^2}{((E^{1\dagger }_k)^T\rho^T_1(E^{1}_k)^T)} = \rho'_1.
\end{aligned}  
\end{equation}
\end{comment}

Finally we can show that $\rho'_1\geq 0$. Denote $\gamma_k = E^1_k\rho_1E^{1\dagger}_k$. Since $\rho_1\geq 0$, we know by the property of positive definite matrix \cite{lax2007linear_algebra} that $\gamma_k$ are all positive definite matrices for all $k$. Again by the property of closure under additivity we conclude that $\rho'_1 = \sum_{k=1}^{d^2}{\gamma_k}$ is a positive definite operator.  

Summarizing claim 1 and claim 2, we conclude that it is equivalent to characterize the attacker's action by $E^1\in B(\mc{H})$ or by $\rho'_1\in \mc{S}$. Similar arguments can be made regarding $E^0\in B(\mc{H})$ and $\rho'_0\in\mc{S}$. This concludes the proof of lemma \ref{lemma:equiv_attacker_action}. 
\end{proof}

\subsection{Proof of Proposition \ref{attacker_strategies_stackelberg_prop}}
\begin{proof}
    We obtain the attacker's optimal strategies by solving the optimization problem \eqref{attacker_stackelberg}. First of all, we can make sure that the optimal strategies $\rho'^*_1,\rho'^*_0$ exist. Since the objective function $u_A$ is convex in terms of $\rho_1$ and $\rho_0$, We apply first-order conditions by taking partial derivatives of $u_A$ in terms of $\rho'_0,\rho'_1$ and set them to be zero, respectively:
\begin{align}
    0 &\equiv \frac{\partial u_A}{\partial \rho'_1} = \Pi^{*}_1 + \lambda (\ln \rho'_1 - \ln \rho_1),
    \label{rho1_first_order_condition}
    \\
      0 &\equiv \frac{\partial u_A}{\partial \rho'_0} =   \ln \rho'_0 - \ln \rho_0, 
    \label{rho0_first_order_condition}
\end{align}
with equality and inequality constraints, which are requirements for $\rho'^*_1,\rho'^*_0$ being density operators, as follows: 
\begin{equation}
\begin{aligned}
    &\tr(\rho'_1) = 1,\;\tr(\rho'_0) = 0, \\
    &\rho'_1 = \rho'^T_1,\;\rho'_0 = \rho'^T_0, \\
    & \rho'_1 \geq 0,\;\rho'_0\geq 0.
\end{aligned}
\end{equation}
By solving \eqref{rho1_first_order_condition} and \eqref{rho0_first_order_condition} we obtain
\begin{equation}
    \begin{aligned}
       \rho'^*_1 & = \frac{1}{Z_1}\exp(\ln\rho_1 - \frac{1}{\lambda}\Pi^*_1), \\
       \rho'^*_0 & = \frac{1}{Z_0}\rho_0,
    \end{aligned}
\end{equation}
where $Z_1,Z_0$ are normalization constants. Referring to the equality constraints we conclude that $Z_0 = 1$ and $Z_1 = \tr(\exp(\ln\rho_1 - \frac{1}{\lambda}\Pi^*_1))$ and arrive at the the solution in \eqref{sol0_attacker_Stackelberg}\eqref{sol1_attacker_Stackelberg}. 
\end{proof}

\subsection{Proof of Proposition \ref{prop:fundamental_limit_detection_rate}}
We make the following assumptions: 
\begin{assume}
We assume the following conditions in proposition 2: 
\begin{enumerate}
    \item $r^1_i>0$ and arranges in a strictly descending order (every eigenvalue is algebraically simple) in terms of $i$;
    \item The basis $\{\keti\}$ is orthonormal; 
    \item The eigenstates $\ketetaa$ are orthonormal; The eigenstates $\ketetaa$ are also arranged in a descending order regarding the eigenvalues of $\rho_1-\tau\rho_0$; The eigenvalues $\eta_j$ are positive whenever $j\geq k_0$;
    \item The difference between eigenvalues are much larger than the projection:
    \begin{equation}
    \forall j\neq i,\;\;\sum\Big|\frac{\brai\Pi^*_1\ketj}{r^1_i-r^1_j}\Big| <1
    \end{equation}
\end{enumerate} 
\end{assume}
The assumptions above make sense since $\rho_1$ is positive definite, symmetric operator. Also, the detector's optimal strategy $\Pi^*_1$ is a projection operator with finite rank.  
We can now state the proof.

\begin{proof}
We adopt the classic perturbation theory for symmetric finite-dimensional linear operators \cite{kato2013perturbation}. Let $|\alpha_j\rangle$ be the eigenstates of the operator $\ln\rho_1-\frac{1}{\lambda}\Pi^*_1$ with eigenvalue $e^{\alpha_j}$. Then we can write 
\begin{equation}
    \exp\left(\ln\rho_1-\frac{1}{\lambda}\Pi^*_1\right) = \sum_{j}{e^{\alpha_j}|\alpha_j\rangle\langle\alpha_j|}.
\end{equation}
Therefore 
\begin{equation}
\begin{aligned}
    &\bar{P}_D(\Pi^*_1) = \tr(\Pi^*_1\rho'^*_1) \\
    &=\frac{1}{Z_1}\tr\Big(\sum_{k\geq k_0}{\ketetaaa\braetaaa}\sum_{j}{e^{\alpha_j}|\alpha_j\rangle\langle\alpha_j|}\Big) \\
    & = \frac{1}{Z_1}\tr\Big(\sum_{k\geq k_0}{\ketetaaa\braetaaa}\sum_{i}{\keteta\braeta} \sum_{j}{e^{\alpha_j}|\alpha_j\rangle\langle\alpha_j|}\sum_{i'}{\ketetapp\braetapp}\Big) \\
    &=\frac{1}{Z_1}\tr\Big(\sum_{k\geq k_0,i}{\ketetaaa\Big(\sum_{j}{e^{\alpha_j}|\alpha_j\rangle\langle\alpha_j|\eta_{i'}\rangle\Big)\braetapp}}\Big) \\
    & = \sum_{j}{\frac{e^{\alpha_j}}{\sum_{j'}{e^{\alpha_{j'}}}}\sum_{k\geq k_0}{|\langle\alpha_j|\eta_{k}\rangle}|^2}
        \label{eq:PD_bar_proof}
\end{aligned}
\end{equation}
On the other hand, we know
\begin{equation}
\begin{aligned}
   &P_D(\Pi^*_1) 
   = \tr(\Pi^*_1\rho_1) \\
   &= \tr\Big(\sum_{k\geq k_0}{\ketetaaa\braetaaa}\sum_{j}{e^{\alpha_j}|\alpha_j\rangle\langle\alpha_j|}\Big) \\
    & =\tr\Big( \sum_{k\geq k_0}{\ketetaaa\braetaaa}\sum_{i}{\keteta\braeta} \sum_{j}{r^1_j\ketj\braj}\sum_{i'}{\ketetapp\braetapp}\Big) \\
    &=\tr\Big(\sum_{k\geq k_0,i}{\ketetaaa\Big(\sum_{j}{e^{\alpha_j}|\alpha_j\rangle\langle\alpha_j|\eta_{i'}\rangle\Big)\braetapp}}\Big) \\
    & = \sum_{j}{r^1_j\sum_{k\geq k_0}{|\braj\eta_{k}\rangle}|^2}.
    \label{eq:PD_proof}
\end{aligned}
\end{equation}
Using the theories of perturbation, consider $\ln\rho_1$ as the unperturbed operator, $\Pi^*_1$ as the perturbation, with $\frac{1}{\lambda}$ controlling the amplitude of the perturbation. Then the eigenvectors, under assumption 1, can be written as a series of $\frac{1}{\lambda}$ as follows:
\begin{equation}
    |\alpha_j\rangle = |\varphi_j\rangle - \frac{1}{\lambda}\sum_{k'\neq j}{\frac{\langle \varphi_{k'}|\Pi^*_1|\varphi_j\rangle}{r^1_{k'}-r^1_j}|\varphi_{k'}\rangle} + o\left(\frac{1}{\lambda^2}\right)
\end{equation}
As a consequence, 
\begin{equation}
\begin{aligned}
  & |\langle \eta_{k} |\alpha_j\rangle|^2 \\
  &= |\langle\eta_{k}|\varphi_j\rangle|^2 - \frac{2}{\lambda}\braetaaa\sum_{k'\neq j}{\frac{\langle \varphi_{k'}|\Pi^*_1|\varphi_j\rangle}{r^1_{k'}-r^1_j}|\varphi_{k'}\rangle}\langle\eta_{k}|\varphi_j\rangle \\
  &+ o\left(\frac{1}{\lambda^2}\right)\leq |\langle\eta_{k}|\varphi_j\rangle|^2.
\end{aligned}
\end{equation}
We can also write out the perturbed eigenvalue $\alpha_j$ in terms of the unperturbed eigenvalue $r^1_j$ as well as the series of the scale $\frac{1}{\lambda}$ as follows:
\begin{equation}
    \alpha_j = \ln r^1_j - \frac{1}{\lambda}\braj\Pi^*_1\ketj + o\left(\frac{1}{\lambda^2}\right).
\end{equation}
As a result,
\begin{equation}
    e^{\alpha_j} = r^1_j\exp\left(-\frac{1}{\lambda}\braj\Pi^*_1\ketj+o\left(\frac{1}{\lambda^2}\right)\right)
\end{equation}
For sufficiently small $1/\lambda$ the higher order term $o(\frac{1}{\lambda^2})$ vanishes.  We get that 
for all $k\geq k_0$ and $j=1,2,\dots, d$,
\begin{equation}
\begin{aligned}
    &\frac{e^{\alpha_j}}{\sum_{j'}{e^{\alpha_{j'}}}}|\langle\alpha_j|\eta_{k}\rangle|^2 \\
    &= \frac{r^1_j\exp\left(-\frac{1}{\lambda}\braj\Pi^*_1\ketj\right)}{\sum_{j'}{r^1_{j'}\exp\left(-\frac{1}{\lambda}\langle\varphi_{j'}|\Pi^*_1|\varphi_{j'}\rangle\right)}}\\
    &\left(|\langle\eta_{k}|\varphi_j\rangle|^2 - \frac{2}{\lambda}\braetaaa\sum_{k'\neq j}{\frac{\langle \varphi_{k'}|\Pi^*_1|\varphi_j\rangle}{r^1_{k'}-r^1_j}|\varphi_{k'}\rangle}\langle\eta_{k}|\varphi_j\rangle\right)
\end{aligned}
\end{equation}
Denote
\begin{equation}
\begin{aligned}
     s^1_j &=\frac{r^1_j\exp\left(-\frac{1}{\lambda}\braj\Pi^*_1\ketj\right)}{\sum_{j'}{r^1_{j'}\exp\left(-\frac{1}{\lambda}\langle\varphi_{j'}|\Pi^*_1|\varphi_{j'}\rangle\right)}} \\ &=\frac{r^1_j\exp\left(-\frac{1}{\lambda}\sum_{k'\geq k_0}{|\braj\eta_{k'}\rangle|^2}\right)}{\sum_{j'}{r^1_{j'}\exp\left(-\frac{1}{\lambda}\sum_{k'\geq k_0}{|\braj\eta_{k'}\rangle|^2}\right)}}.
\end{aligned}
\end{equation}
Noticing that $\sum_{j}{r^1_j} =\sum_{j}{s^1_j}= 1$. Also if for some $j_1,j_2\leq d$, $\sum_{k\geq k_0}{|\braj\eta_{k}\rangle|^2 }\leq \sum_{k\geq k_0}{|\langle\varphi_{j'}|\eta_{k}\rangle|^2 } $ is large, then we must have $s^1_{j'}\geq s^1_{j}$. By generalization of AM-GM inequality we conclude that
\begin{equation}
    \sum_{j}{\sum_{k\geq k_0}{\frac{e^{\alpha_j}}{\sum_{j'}{e^{\alpha_{j'}}}}|\langle\alpha_j|\eta_{k}\rangle|^2}} \leq \sum_{j}{\sum_{k\geq k_0}{r^1_j|\braj\eta_{k}\rangle|^2 }}.
\end{equation}
Noticing the expression of $P_D(\Pi^*_1)$ in \eqref{eq:PD_proof} and $\bar{P}_D(\Pi^*_1)$ in \eqref{eq:PD_bar_proof}, we get
\begin{equation}
      \bar{P}_D(\Pi^*_1)\leq P_D(\Pi^*_1).
\end{equation}
On the other hand, $s^1_j\geq \frac{r^1_je^{-\frac{1}{\lambda}}}{1- \frac{1}{\lambda} + \frac{1}{\lambda^2}}$ for all $j$. We also notice 
\begin{equation}
    |\langle\alpha_j|\eta_{k}\rangle|^2\geq (1-\frac{1}{\lam})|\langle\varphi_j|\eta_{k}\rangle|^2,\;\forall j,k\geq k_0.
\end{equation}
As a result, we have
\begin{equation}
     \bar{P}_D(\Pi^*_1)\geq P_D(\Pi^*_1)e^{-\frac{1}{\lambda}},
\end{equation}
which concludes the proof. 
\end{proof}

\end{document}